\documentclass[twocolumn,showpacs,preprintnumbers,amsmath,amssymb]{revtex4}

\usepackage{graphicx}
\usepackage{dcolumn}
\usepackage{bm}

\begin{document}


%
\title{Band-Selective Modification of the Magnetic Fluctuations in 
Sr$_{2}$RuO$_{4}$: \\ 
Study of Substitution Effects
}

%
\author{Naoki Kikugawa}
\affiliation{
School of Physics and Astronomy, University of St.~Andrews, 
St.~Andrews, Fife KY16 9SS, United Kingdom\\
Venture Business Laboratory, Kyoto University, 
Kyoto 606-8501, Japan\\
Department of Physics, Kyoto University, Kyoto 606-8502, Japan
}

\author{Christoph Bergemann}
\affiliation{
Cavendish Laboratory, University of Cambridge, 
Madingley Road, Cambridge CB3 0HE, United Kingdom
}

\author{Andrew Peter Mackenzie}
\affiliation{
School of Physics and Astronomy, University of St.~Andrews, 
St.~Andrews, Fife KY16 9SS, United Kingdom
} 

\author{Yoshiteru Maeno}
\affiliation{
International Innovation Center, Kyoto University, 
Kyoto 606-8501, Japan\\
Department of Physics, Kyoto University, Kyoto 606-8502, Japan
}

%
\date{\today}

%
\begin{abstract}
We report a study of magnetic, thermal, and transport properties of 
La$^{3+}$ substituted Sr$_{2}$RuO$_{4}$, 
performed in order to investigate the effects of additional electron doping 
in this correlated metal. 
A gradual enhancement of the electronic part of specific heat and a more drastic
increase of the static magnetic susceptibility were observed 
in Sr$_{2-y}$La$_{y}$RuO$_{4}$ with increasing $y$. 
Furthermore, 
the quasi-two-dimensional Fermi-liquid behavior 
seen in pure Sr$_{2}$RuO$_{4}$ breaks down 
near the critical concentration $y_{\rm c}$\,$\sim$\,0.20.
Combined with a realistic tight-binding model 
with rigid-band shift of Fermi level, 
the enhancement of the density of states can be ascribed 
to the elevation of the Fermi energy toward a van Hove singularity of 
the thermodynamically dominant $\gamma$ Fermi-surface sheet.
On approaching the van Hove singularity, 
the effective nesting-vector of the $\gamma$ band shrinks 
and further enhances the susceptibility near the wave vector $q$\,$\sim$\,0. 
We attribute the non-Fermi-liquid behavior to two-dimensional ferromagnetic fluctuations 
with short range correlations at the van Hove singularity.
The observed behavior is in sharp contrast to that of 
Ti$^{4+}$ substitution in Sr$_{2}$RuO$_{4}$ 
which enhances antiferromagnetic fluctuations and 
subsequently induces incommensurate magnetic ordering 
associated with the nesting between the other Fermi-surface sheets 
($\alpha$ and $\beta$).
We thus establish that 
substitution of appropriate chemical dopants can band-selectively
modify the spin-fluctuation spectrum in the spin-triplet superconductor Sr$_{2}$RuO$_{4}$.
\end{abstract}


\maketitle

%
\section{Introduction}

Since the discovery of the superconductivity in the layered perovskite Sr$_{2}$RuO$_{4}$ \cite{Maeno_Nature}, 
the material has been the subject of intense research \cite{Andy_Review} 
for the following reasons.
First, 
the superconductivity (with transition temperature $T_{\rm c}$\,=\,1.5\,K) is 
of unconventional pairing symmetry, 
most probably spin triplet \cite{Andy_Review}. 
Second, 
a highly conductive metallic state with mean free path 
$\ell$\,$>$\,1\,$\mu$m can be achieved, 
reflecting the well hybridized Ru 4$d$ and O 2$p$ character 
of the conduction bands in the stoichiometric material. 
This made it possible to clarify the detailed electronic structure 
by means of de Haas-van Alphen (dHvA) experiments \cite{Andy_dHvA,Bergemann_PRL,Bergemann_Review} 
and angle-resolved photoemission-spectra (ARPES) 
with \cite{Damascelli_ARPES} results that 
are in qualitative agreement with band structure calculations \cite{Oguchi,Singh_Band}: 
the Fermi surface consists of one hole sheet ($\alpha$) and two electron sheets ($\beta$ and $\gamma$). 
The $\alpha$ and $\beta$ bands are formed by the Ru $d_{yz}$ and 
$d_{zx}$ orbits, while the $\gamma$ band has $d_{xy}$ orbital character. 
On the basis of the cylindrical Fermi-surface topography, 
normal-state properties are described 
quantitatively within the framework of a quasi-two-dimensional 
Fermi-liquid \cite{Maeno_FL,Bergemann_Review}. 
For a strongly correlated, 
unconventional superconductor like Sr$_{2}$RuO$_{4}$, 
knowledge of the relationship 
between the superconductivity and the magnetic fluctuations is 
of fundamental importance in order to clarify the pairing mechanism. 
This is strikingly exemplified by a number of $f$-electron systems 
and in the high-$T_{\rm c}$ cuprates, 
where the superconductivity emerges 
near regions of magnetic instability \cite{Mathur,Saxena,Aeppli}.
In those cases, 
the magnetic instability point is reached by driving the magnetic ordering temperature to zero by tuning parameters 
such as pressure and carrier doping by chemical substitution 
\cite{Stewart_Review}. 
Also, 
non-Fermi-liquid normal state behavior is often observed 
in the vicinity of the quantum critical points. 
Spin fluctuations are thought to act as the ``magnetic glue'' 
responsible for the superconducting pairing in many of these cases,  
including, as theoretically suggested, Sr$_{2}$RuO$_{4}$ 
\cite{Monthoux}. 
Experimentally, 
a more direct signature of spin fluctuations can be found 
in inelastic neutron scattering measurements. 
In Sr$_{2}$RuO$_{4}$, 
recent experiments revealed 
a weak, broadened structure around the wave vector
$\textit{\textbf{q}}$\,$\sim$\,$\textbf{0}$, 
attributed to the excitation from the $\gamma$ band \cite{Braden_Sr2RuO4}, 
in addition to a well-known feature at the incommensurate wave vector
$\textit{\textbf{q}}$\,=\,$\textit{\textbf{Q}}_{\rm ic}^{{\alpha}{\beta}}$\,$\sim$\,(2$\pi$/3,\,2$\pi$/3,\,0) \cite{Sidis}. 
The incommensurate wave vector is in accord with a nesting vector 
between the $\alpha$ and $\beta$ Fermi surfaces \cite{Mazin_PRL99}. 
Attempts have been made to probe the proximity 
to magnetic quantum critical points 
by applying strong magnetic field \cite{Bergemann_HighField} 
or hydrostatic pressure \cite{Forsythe}. 
These probes have the advantage of being ``clean'', i.e. 
of not introducing disorder, 
but in the range of parameters used so far on Sr$_{2}$RuO$_{4}$, 
no magnetic instabilities have been discovered. 
Although substitution studies involve the introduction of disorder, they have the significant advantage of producing band-specific effects.  For example,
the substitution of nonmagnetic Ti$^{4+}$ (3$d^{0}$) for Ru$^{4+}$ 
(4$d^{4}$) in Sr$_{2}$Ru$_{1-x}$Ti$_x$O$_{4}$ is a powerful probe 
to enhance \textit{only} the anisotropic antiferromagnetic fluctuations 
at $\textit{\textbf{Q}}_{\rm ic}^{{\alpha}{\beta}}$ 
\cite{Braden_Ti,Kiku_PRL,Ishida_Ti}: 
the ground state in Sr$_{2}$Ru$_{1-x}$Ti$_{x}$O$_{4}$ changes 
from spin-triplet superconductivity 
to incommensurate spin-density-wave (SDW) order \cite{Braden_Ti} 
with the formation of glassy clusters at lower temperatures 
\cite{Minakata,Klaus}. 
Near the onset of magnetic order at $x$\,=\,$x_{\rm c}$\,$\sim$\,0.025, 
breakdown of the Fermi-liquid behavior is observed at low temperatures: 
the resistivity and specific heat show linear and 
logarithmic temperature dependence respectively \cite{Kiku_PRL}. 
These results indicate that in Sr$_{2}$Ru$_{1-x}$Ti$_{x}$O$_{4}$ 
the divergence of spin fluctuations at 
$\textit{\textbf{Q}}_{\rm ic}^{{\alpha}{\beta}}$, 
arising from the nesting between $\alpha$ and $\beta$ bands, 
dominates the behavior at $x_{\rm c}$. 
%

\begin{figure}
\includegraphics[width=70mm]{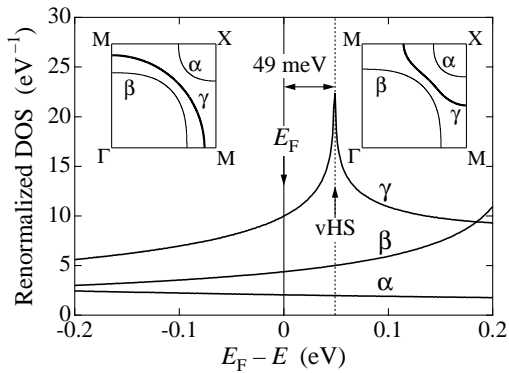}
\caption{\label{Fig1} 
Partial density of states (DOS)  in Sr$_{2}$RuO$_{4}$ 
obtained by a tight-binding fit to the experimentally determined 
Fermi surface \cite{Bergemann_Review}. 
The Fermi energy is located 49\,meV 
below a van Hove Singularity (vHS) of the $\gamma$ band. 
Note that the $\gamma$ Fermi-surface changes 
from an electron pocket (left inset) to a hole pocket (right inset) 
when the Fermi energy crosses the van Hove singularity 
at additional electron doping level of $y_{\rm c}$\,$\sim$\,0.23/f.u. 
}
\end{figure}

%
On the other hand, 
the role of the magnetic fluctuations of the $\gamma$ band, 
which is proposed by some authors to drive 
the superconductivity in Sr$_{2}$RuO$_{4}$ \cite{Sigrist,Nomura}, 
still remains unclear. 
The $\gamma$ band has the largest density of states (DOS) 
at the Fermi energy and the strongest mass enhancement 
\cite{Bergemann_Review}. 
As shown in Figure~\ref{Fig1}, 
in a tight-binding fit to the experimental Fermi-surface geometry, 
the Fermi energy is located 49\,meV 
below a van Hove singularity (vHS) of the $\gamma$ band \cite{Bergemann_Review}, 
corresponding to electron doping of an additional 0.23 electrons 
per formula unit 
\cite{Bergemann_Review,Electron_Comment}. 
In this paper, 
we report the effect of non-isovalent counter-ion 
substitution of Sr$^{2+}$ with La$^{3+}$, 
as in Sr$_{2-y}$La$_{y}$RuO$_{4}$. 
The primary effect of La doping is the introduction of 
additional electrons to the metallic bands at the Fermi energy. 
Also, 
structural distortions are minimized because of the similar ionic radii 
between Sr$^{2+}$ and La$^{3+}$. 
La substitution therefore provides a gentle way 
to study the effect of changing carrier concentration 
in the correlated metal and unconventional superconductor 
Sr$_{2}$RuO$_{4}$. 
We achieved electron doping up to $y$\,=\,0.27, 
where the tight-binding calculation places the Fermi energy of the material 
well $beyond$ the vHS. 
The main contribution of the enhancement of the DOS is confirmed 
to be due to the approach towards the vHS of the $\gamma$ band. 
At the same time, 
the nesting wave vector of the $\gamma$ band is 
shrinking towards $q$\,$\sim$\,0 at the vHS, 
further enhancing the low-$q$ susceptibility. 
We observe non-Fermi-liquid behavior 
around the ``critical'' doping level of $y_{\rm c}$\,$\sim$\,0.20 
and attribute it to two-dimensional ferromagnetic fluctuations 
with short range correlations. 
The evolution of the ferromagnetic fluctuations with electron doping is 
in sharp contrast to the enhancement of the antiferromagnetic fluctuations 
induced by Ti substitution \cite{Kiku_PRL,Ishida_Ti}. 
Throughout this paper, 
we stress that substitution of appropriate dopants into Sr$_{2}$RuO$_{4}$ 
can band-selectively modify the magnetic fluctuation spectrum. 
%

%
\section{Experimental}

Single crystals of Sr$_{2-y}$La$_{y}$RuO$_{4}$  with $y$ up to 0.27 were 
grown by a floating-zone method in an infrared image furnace 
(NEC Machinery, model SC-E15HD). 
Although it was difficult to grow crystals with increasing $y$ 
because of the necessity of higher temperature and 
therefore an unstable molten zone during crystal growth, 
we finally succeeded in obtaining large crystals with typical size of 
4\,mm\,$\times$\,3\,mm\,($c$ axis)\,$\times$\,60\,mm. 
The La concentrations of the crystals were analyzed by 
electron-probe microanalysis (EPMA). 
Up to $y$\,= \,0.14, the La substitutes well for Sr. 
On the other hand, 
we found that the analyzed La concentration $y_{\rm a}$ 
deviates from the nominal La concentration 
$y_{\rm n}$ for $y_{\rm n}$\,$>$\,0.14: 
the $y_{\rm a}$ varies roughly as $y_{\rm a}$\,$\sim$\,0.3$y_{\rm n}$\,+\,0.12.
We confirmed tetragonal crystal symmetry 
for all Sr$_{2-y}$La$_{y}$RuO$_{4}$ crystals used in this study 
at room temperature from X-ray measurement. 
The lattice parameter along the in-plane ($a$ axis) 
increases by $\sim$\,0.5\% and 
that perpendicular to the plane ($c$ axis) decreases 
by $\sim$\,0.4\% continuously with $y$ up to $y$\,=\,0.27. 
The crystal symmetry and the change of lattice parameter by La doping are 
consistent with previous work on polycrystals \cite{Hashimoto}. 
Magnetic susceptibility measurements were performed using 
a commercial superconducting quantum interference device 
magnetometer down to 1.8\,K (Quantum Design, MPMS-XL). 
The specific heat $C_{P}$ was measured by a thermal relaxation method
from 0.5\,K to 30\,K (Quantum Design, model PPMS). 
The electrical resistivity was measured by standard four-probe dc and 
ac methods between 4.2 and 300\,K and by an ac method between 
0.3 and 5\,K.
%

%
\section{Results}

\subsection{Static magnetic susceptibility}

\begin{figure}
\includegraphics[width=70mm]{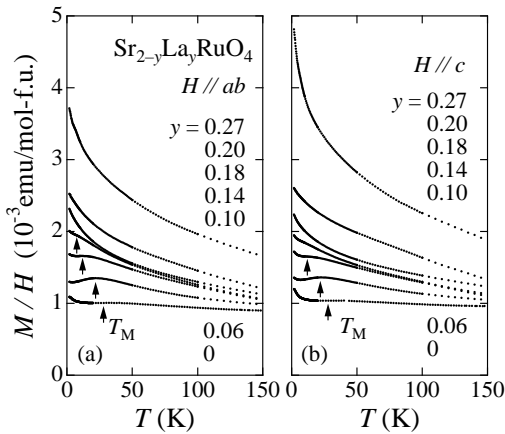}
\caption{\label{Fig2} 
Temperature dependence of the magnetic susceptibility 
$\chi$($T$)\,=\,$M/H$ of Sr$_{2-y}$La$_{y}$RuO$_{4}$ 
with $y$ up to 0.27 in applied field of 1\,T 
along (a) the basal ($ab$) plane and (b) the $c$ axis. 
$T_{\rm M}$ indicates the characteristic temperature at which $M/H$ 
shows a peak. 
}
\end{figure}

%
Figure~\ref{Fig2} shows 
the temperature dependence of the static magnetic susceptibility 
$\chi$($T$)\,=\,$M/H$ of Sr$_{2-y}$La$_{y}$RuO$_{4}$ with $y$ up to 0.27 
in an applied field of 1\,T along the basal ($ab$) plane (a) and the $c$ axis (b). 
There is no clear sign of any magnetic ordering 
for all dopant concentrations in this study \cite{Order_Comment}.
It should be noted that the magnetic susceptibility is nearly 
\textit{isotropic} with respect to the direction of the applied magnetic field. 
This behavior is in sharp contrast with that of Ti-substituted Sr$_{2}$RuO$_{4}$, 
which exhibits Ising anisotropy \cite{Minakata,Klaus}. 
In the normal state of pure Sr$_{2}$RuO$_{4}$, 
$\chi$($T$) shows little temperature dependence (Pauli paramagnetism) 
with a broad peak around 30\,K (denoted as $T_{\rm M}$), 
coinciding with the temperature
below which the characteristic signatures of a Fermi-liquid are seen. 
With increasing $y$, 
a gradual change to Curie-Weiss-like behavior occurs at high temperature.
For $y$\,=\,0.06, 
the behavior with the peak temperature $T_{\rm M}$\,$\sim$\,22\,K 
is similar to that seen in the enhanced paramagnet 
Sr$_{3}$Ru$_{2}$O$_{7}$ 
which possesses a ferromagnetic instability under uniaxial stress \cite{Ikeda}. 
$T_{\rm M}$ shifts to lower temperature with further La doping, 
and finally for $y$\,$>$\,0.18 the susceptibility continues to rise sharply 
down to the lowest temperatures. 
The effective magnetic moment $p_{\rm eff}$, 
as well as the temperature independent term $\chi_{\rm Pauli}$ 
corresponding to the Pauli paramagnetism, 
estimated from the Curie-Weiss fitting at higher temperature, 
increase linearly at the rate of 
$dp_{\rm eff}$/$dy$\,$\sim$\,2$\mu_{\rm B}$/La and 
$d\chi_{\rm Pauli}$/$dy$\,$\sim$\,0.7$\times$10$^{-3}$\,(emu/mol-f.u.)/La, respectively. 
%

\subsection{Specific heat}

\begin{figure}
\includegraphics[width=70mm]{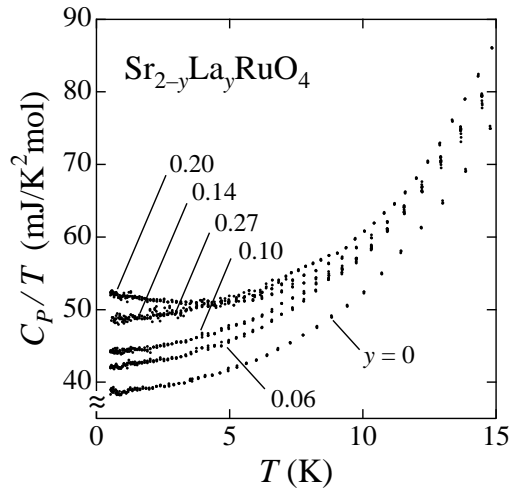}
\caption{\label{Fig3} 
Temperature dependence of $C_{P}/T$ in Sr$_{2-y}$La$_{y}$RuO$_{4}$ 
up to $y$\,=\,0.27. 
For pure Sr$_{2}$RuO$_{4}$ ($y$\,=\,0) a magnetic field of 0.2\,T 
was applied along the $c$ axis to suppress the superconductivity. 
}
\end{figure}

%
Figure~\ref{Fig3} shows the temperature dependence of specific heat 
divided by temperature $C_{P}/T$ for Sr$_{2-y}$La$_{y}$RuO$_{4}$ 
up to $y$\,=\,0.27 and down to $T$\,=\,0.5K.
There is no sign of a phase transition, magnetic or otherwise, 
in line with the susceptibility results. 
The data for pure Sr$_{2}$RuO$_{4}$ ($y$\,=\,0) were obtained by 
applying a magnetic field of 0.2\,T along the $c$ axis 
in order to suppress the superconductivity ($T_{\rm c}$\,=\,1.44\,K); 
the data for $\mu_{0}H$\,=\,0\,T essentially overlap with the data for 0.2\,T 
for $T$\,$>$\,1.5\,K \cite{Andy_JPSJ}. 
Enhancement of $C_{P}/T$ with $y$ is clearly observed and 
is due to the increase of the DOS by electron doping. 
The enhancement agrees well with the calculations 
using tight-binding parameters 
based on the rigid-band model that each La dopes one free-electron 
\cite{Kiku_Rigid}. 
Moreover, 
a clear low-temperature upturn indicating a deviation 
from simple Fermi-liquid behavior is observed around $y$\,=\,0.20. 
This upturn can not simply be explained by an impurity effect 
induced by La doping, 
because the effect is suppressed for $y$\,=\,0.27. 
This behavior is similar to that seen in Ti-substituted Sr$_{2}$RuO$_{4}$ 
in the vicinity of its magnetic ordering \cite{Kiku_PRL}.
%

\subsection{In- and out-of-plane resistivity}

\begin{figure}
\includegraphics[width=70mm]{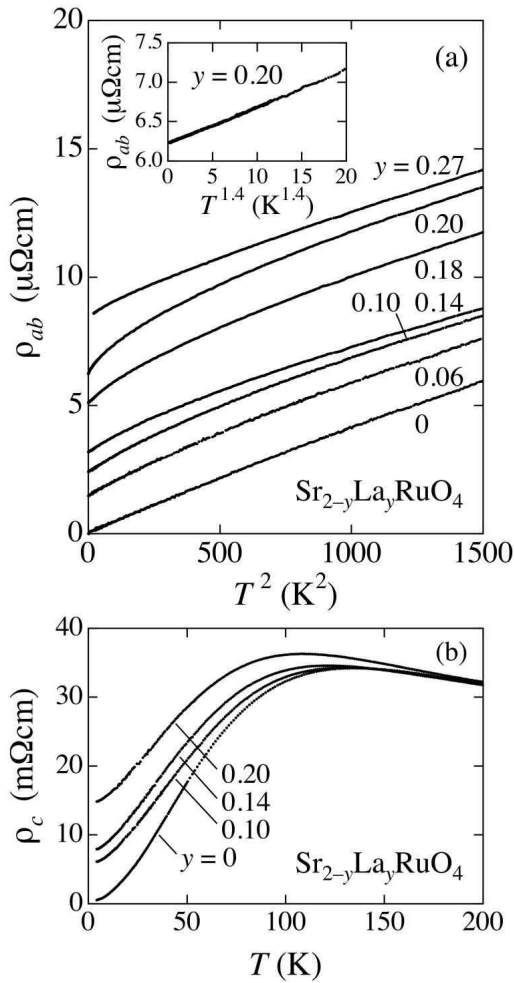}
\caption{\label{Fig4} 
(a) In-plane resistivity $\rho_{ab}$ plotted against $T^{2}$ 
in Sr$_{2-y}$La$_{y}$RuO$_{4}$. 
The inset shows $\rho_{ab}$ vs. $y$ with $y$\,=\,0.20. 
(b) Temperature dependence of the $c$-axis resistivity 
in Sr$_{2-y}$La$_{y}$RuO$_{4}$ with several $y$. 
}
\end{figure}

%
Figure~\ref{Fig4}(a) shows the in-plane resistivity $\rho_{ab}$ 
for various doping levels $y$, plotted against $T^{2}$. 
The residual resistivity $\rho_{ab0}$, 
defined by the extrapolation of the low temperature resistivity to $T$\,=\,0\,K, increases systematically with $y$ 
at the rate of $d\rho_{ab0}/dy$\,$\sim$\,40\,$\mu\Omega$cm, 
that is, 
with a phase shift of impurity scattering $\delta_{0}$\,$\sim$\,$\pi$/12. 
The enhancement is much smaller than that seen in in-plane substituted
Sr$_{2}$RuO$_{4}$ such as Sr$_{2}$Ru$_{1-x}$Ti$_{x}$O$_{4}$ and
Sr$_{2}$Ru$_{1-x}$Ir$_{x}$O$_{4}$, 
where both Ti$^{4+}$ and Ir$^{4+}$ impurities act as 
unitary potential scatterers with $\delta_{0}$\,$\sim$\,$\pi/2$ \cite{Kiku_JPSJ}. 
This result shows that out-of-plane La$^{3+}$ substitution 
for Sr$^{2+}$ introduces less severe disorder 
within the conductive RuO$_{2}$ planes. 
Another important result in Fig.~\ref{Fig4}(a) is 
the breakdown of the Fermi-liquid behavior around $y$\,=\,0.20: 
the $T$-squared dependence 
$\rho_{ab}$\,=\,$AT^{n}$\,+\,$\rho_{ab0}$ with $n$\,=\,2, 
satisfied below about 30\,K for pure Sr$_{2}$RuO$_{4}$ \cite{Maeno_FL}, 
starts to break down with $y$. 
This is displayed in Fig.~\ref{Fig4}(a) by the fact that 
the temperature range in which the $T$-squared fitting is valid 
shrinks with $y$ \cite{Adachi_Comment}. 
Also, 
the coefficient $A$\,$\propto$\,$m_{\rm e}^{2}$/$n_{\rm e}$ 
gradually increases around $y$\,$\sim$\,0.20, 
indicating the enhancement of the correlation among electrons 
by La doping. 
Here, 
$m_{\rm e}$ and $n_{\rm e}$ are the electron effective mass and 
carrier density, respectively. 
Our best fit for $y$\,=\,0.20 is obtained with $n$\,=\,1.4\,$\pm$\,0.05. 
As added in the inset of Fig.~\ref{Fig4}, 
the $T^{1.4}$ behavior is well satisfied between 0.3 and 10\,K 
over more than one decade in temperature. 
It should be noted that such a deviation from Fermi-liquid behavior is
not explained by an effect of disorder by La, 
because the $T$-squared behavior is recovered for $y$\,=\,0.27 
below $\sim$\,7\,K.
The temperature dependence of the resistivity along the $c$ axis,
$\rho_{c}$, for various $y$ is presented in Figure~\ref{Fig4}(b).
The crossover temperature from non-metallic behavior 
($d\rho_{c}/dT$\,$<$\,0) at high temperature 
to metallic one ($d\rho_{c}/dT$\,$>$\,0) at low temperature \cite{Yoshida} 
moves monotonically to lower temperature with $y$ 
as seen in Ti-substituted Sr$_{2}$RuO$_{4}$ \cite{Minakata}. 
The residual resistivity $\rho_{c0}$ increases with doping 
with a slope $d\rho_{c0}/dy$\,$\sim$\,70\,m$\Omega$\,cm/$y$. 
At the same time, 
the resistivity anisotropy $\rho_{c}$/$\rho_{ab}$ 
at low temperatures remains 
around 2\,$\times$\,10$^{3}$ for all dopant concentrations. 
Also, 
the temperature region over which a Fermi-liquid-like $T^2$ law
is valid for $\rho_c$ is almost identical to that for $\rho_{ab}$.
This result implies that 
the low-temperature transport mechanism
along the $c$ axis is hardly affected by electron doping: 
the quasi-particles can propagate coherently between the RuO$_{2}$ layers 
at low temperatures. 
%

\subsection{Phase diagram of Sr$_{2-y}$La$_{y}$RuO$_{4}$}

\begin{figure}
\includegraphics[width=70mm]{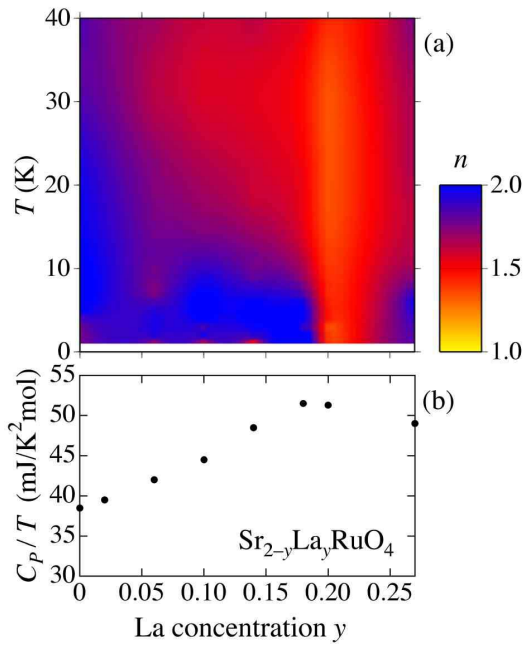}
\caption{\label{Fig5} 
(a) Temperature and La concentration evolution of the exponent $n$ 
in Sr$_{2-y}$La$_{y}$RuO$_{4}$. 
The exponent $n$ is derived from the expression $\rho_{ab}$\,=$\rho_{ab0}$\,+$AT^{n}$. 
(b) the value of $C_{P}$/$T$ at 0.5\,K as a function of $y$. 
 }
\end{figure}

%
The phase diagram of 
Sr$_{2-y}$La$_{y}$RuO$_{4}$ is presented in Fig.~\ref{Fig5} 
from (a) the in-plane resistivity and (b) the specific heat measurements. 
The substitution of La leads to 
the suppression of the characteristic temperature 
of the Fermi-liquid behavior \cite{Adachi_Comment} 
in addition to the suppression of $T_{\rm c}$ around $y$\,$\sim$\,0.03 
as described in ref.~\onlinecite{Kiku_Rigid}. 
Finally, 
for $y$\,$\sim$\,0.20,  
we can see the breakdown of the Fermi-liquid behavior ($n$\,=\,1.4) 
at low temperature. 
Also, 
the enhancement of $C_{P}/T$ at 0.5\,K in Fig.~\ref{Fig5}(b) 
suggests deviation from Fermi-liquid behavior 
in the temperature dependence of $C_{P}/T$ (Fig.~\ref{Fig3}). 
This behavior may imply the presence of a quantum critical point, 
near $y$\,=\,0.20, 
although no magnetically ordered state has been identified in this study. 
The critical concentration $y$\,$\sim$\,0.20 is in good agreement with 
the prediction from tight-binding calculations, 
where $y_{\rm c}$ is evaluated as 0.23 
\cite{Bergemann_Review,Electron_Comment}. 
%

%
\section{Discussion}

\subsection{Enhancement of the density of states of the $\gamma$ band 
by electron doping in Sr$_{2}$RuO$_{4}$}

Let us first discuss the origin of the enhancement of the density of states 
by electron doping as seen in the specific heat in Fig.~\ref{Fig3}. 
In the band calculation based on a tight-binding model, 
whose parameters are determined by a fit to the experimentally-observed 
Fermi surfaces \cite{Bergemann_Review},  
the Fermi energy is located 49\,meV below a vHS of the $\gamma$ band 
as shown in Fig.~\ref{Fig1}, 
whereas such a singularity is not expected for the $\alpha$ and $\beta$ bands 
\cite{Singh_Band,Bergemann_Review}. 
The Fermi energy coincides with the singularity 
at a doping level of an additional 0.23 electrons/f.u., 
if a rigid-band model is applicable to this system \cite{Bergemann_Review}.
Very recently, 
we confirmed the rigid-band model to be valid 
with good quantitative agreement in Sr$_{2-y}$La$_{y}$RuO$_{4}$ 
up to $y$\,=\,0.06 
in a comparison between quantum oscillation measurements 
(de Haas-van Alphen effect) and tight-binding calculations \cite{Kiku_Rigid}. 
Here we find experimentally that the critical doping value 
appears to be 0.20 rather than 0.23 electrons/f.u. (Fig.~\ref{Fig5}). 
This indicates a very slight departure from the rigid-band-shift model,
possibly because the decrease of the $c$-axis lattice parameter 
with La substitution, which lowers the $\gamma$ band 
relative to the others, might not be negligible for higher La concentration \cite{Kyle_Comment}. 
%

\subsection{Enhancement of the $\gamma$ band magnetic susceptibility
at \textit{q}\,$\sim$\,0 by electron doping in Sr$_{2}$RuO$_{4}$} 

\begin{figure}
\includegraphics[width=70mm]{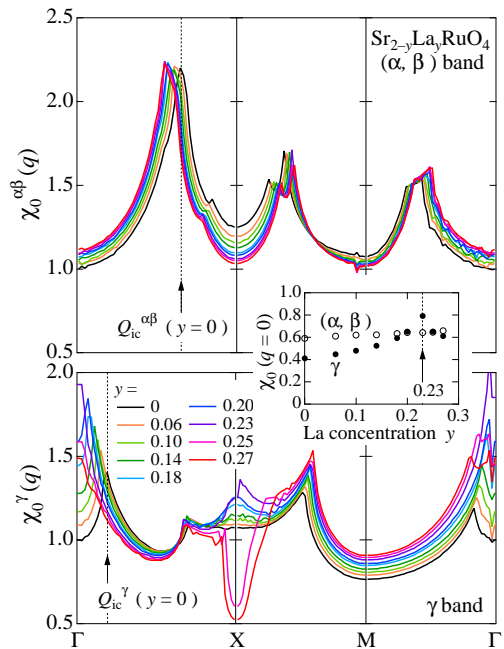}
\caption{\label{Fig6} 
Momentum dependence of real part of Lindhard susceptibility 
of Sr$_{2-y}$La$_{y}$RuO$_{4}$ up to $y$\,=\,0.27 
based on the realistic tight-binding band-structure and a rigid-band shift. 
The contributions of the $\alpha$, $\beta$ bands (upper panel) and 
the $\gamma$ band (lower panel) are presented separately, 
normalized to their values at $q$\,=\,0 and $y$\,=\,0. 
The inset shows the La concentration dependence of 
the static susceptibility ($q$\,=\,0) of 
the the $\alpha$, $\beta$ bands (open circles) and 
the $\gamma$ band (closed circles), 
with the $y$\,=\,0 values indicating the relative contributions 
of ($\alpha$, $\beta$) and $\gamma$ to the bare density of states. 
}
\end{figure}

\begin{figure*}
\includegraphics[width=135mm]{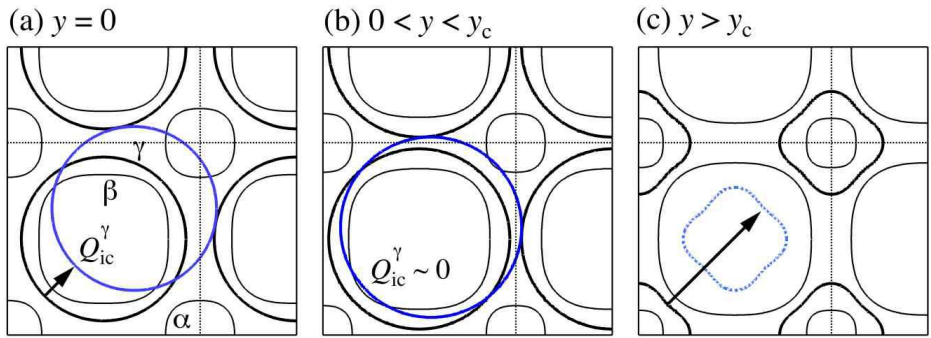}
\caption{\label{Fig7} 
Sketch of the \textit{weak} Fermi-surface nesting 
between $\gamma$ bands for 
(a) $y$\,=\,0, (b) 0\,$<$\,$y$\,$<$\,$y_{\rm c}$, and (c) $y$\,$>$\,$y_{\rm c}$. 
Note that the effective nesting vector 
$\textit{\textbf{Q}}_{\rm ic}^{\gamma}$ shrinks with $y$. 
Also, the nesting vector around ($\pi$,\,$\pi$) for $y$\,$>$\,$y_{\rm c}$ is added 
in order to explain the suppression of the $\chi_{0}^{\gamma}$($q$) near ($\pi$,\,$\pi$). }
\end{figure*}

%
On the basis of the rigid-band model described above, 
we now discuss the contributions from the $\gamma$ band 
to the magnetic and thermal properties 
when additional electrons are doped into Sr$_{2}$RuO$_{4}$. 
Figure~\ref{Fig6} shows the momentum dependence of 
the \textit{real} part of the Lindhard susceptibility $\chi_{0}$($q$) \cite{Yoshida_Textbook} expected for Sr$_{2-y}$La$_{y}$RuO$_{4}$ 
up to $y$\,=\,0.27. 
Again, 
the calculation is performed using a tight-binding fit 
to the experimentally determined Fermi surface geometry 
\cite{Bergemann_Review}, 
assuming a rigid-band shift. 
It should be noted that the Lindhard spin susceptibility calculated
here is \textit{isotropic} although the electronic structure itself is anisotropic, 
because the the susceptibility formula represents 
the Zeeman splitting in a metallic state. 
The figure is separated out into the contributions 
from the $\alpha$, $\beta$ bands (upper panel of Fig.~\ref{Fig6}) 
and $\gamma$ band (lower panel of Fig.~\ref{Fig6}).
The susceptibility is normalized at $q$\,=\,0 in pure Sr$_{2}$RuO$_{4}$, 
and the Stoner factor is neglected in this calculation for the simplicity. 
Comparing the contribution from the $\alpha/\beta$ and 
$\gamma$ bands, 
$\chi_{0}^{\gamma}(q)$ changes much more dramatically on 
electron doping than $\chi_{0}^{{\alpha}{\beta}}$($q$) 
which changes little apart from a small shift in 
$\textit{\textbf{Q}}_{\rm ic}^{{\alpha}{\beta}}$. 
This strongly suggests that 
it is the $\gamma$ band that mostly affects the change 
in both the magnetic and thermal properties by electron doping. 
Especially, 
it is clearly found that $\chi_{0}^{\gamma}$($q$\,=\,0) mainly 
contributes to the enhancement of the total static susceptibility 
by electron doping, 
as shown in the inset of Fig.~\ref{Fig6}. 
The peak in $\chi_0^{\gamma}(q)$ at 
$\textit{\textbf{q}}$\,=\,$\textit{\textbf{Q}}_{\rm ic}^{\gamma}$\,$\sim$\,(0.2$\pi$,\,0.2$\pi$) 
as seen in pure Sr$_{2}$RuO$_{4}$ is explained 
by weak nesting of the two-dimensional $\gamma$ bands 
at the nesting vector $\textit{\textbf{Q}}_{\rm ic}^{\gamma}$, 
as illustrated in Figure~\ref{Fig7}(a). 
A corresponding structure has only recently been observed 
in inelastic neutron scattering measurement \cite{Braden_Sr2RuO4}, 
where the experiments detect the \textit{imaginary} part 
of the dynamical susceptibility. 
The main effect of electron doping is that 
the nesting wave vector $\textit{\textbf{Q}}_{\rm ic}^{\gamma}$ shifts 
to lower $q$, 
while both 
$\chi_{0}^{\gamma}(q\,=\,0)$ and
$\chi_{0}^{\gamma}(\textit{\textbf{Q}}_{\rm ic}^{\gamma})$ diverge.
This divergence of $\chi_{0}^{\gamma}$ is logarithmic in $(y-y_{\rm c})$, 
but one would expect 
the Stoner enhancement to lead to a much more dramatic increase 
in the renormalized susceptibility $\chi^{\gamma}$. 
Finally, 
$\textit{\textbf{Q}}_{\rm ic}^{\gamma}$ becomes zero at the
van Hove singularity itself, 
at a critical electron concentration $y_{\rm c}$\,=\,0.23 (Figure~\ref{Fig7}(b)). 
Beyond the singularity ($y$\,$>$\,$y_{\rm c}$), 
$\chi_{0}^{\gamma}$($q$) at $\textit{\textbf{q}}$\,$\sim$\,($\pi$,\,$\pi$) 
(the X point) is dramatically suppressed, 
as seen for $y$\,=\,0.25 and 0.27 in Figure~\ref{Fig6}(b). 
Here the Fermi energy is higher than the vHS, 
and as illustrated in the right inset of Figure~\ref{Fig1}, 
the Fermi surface of the $\gamma$ band changes 
from an electron pocket to a hole one. 
As seen in Figure~\ref{Fig7}(c), 
the wave vector ($\pi$,\,$\pi$) then fails to connect 
the $\gamma$ sheet with itself, 
leading to a substantial loss of susceptibility near the X point. 
Using the above Lindhard calculation, 
although the Stoner factor is neglected in this calculation, 
we can qualitatively explain most of 
the evolution of the results of our our static susceptibility 
($q$\,=\,0) with La doping (Fig.~\ref{Fig2}). 
The marked increase of the observed susceptibility with La doping 
is explained in terms of the enhanced density of states 
at the Fermi level as the vHS is approached and 
$\textit{\textbf{Q}}_{\rm ic}^{\gamma}$ shrinks to zero. 
We note again that the Lindhard susceptibility is always \textit{isotropic},
in accordance with our experiment (Figure~\ref{Fig2}), 
since it essentially represents the Zeeman splitting in the metallic state. 
This holds despite the quasi-two-dimensionality of the electronic
structure and spin-fluctuation spectrum. 
However, 
we note that 
there must also be another contribution to the susceptibility as well, 
because it is not easy to account in the above analysis 
either for the strength of the temperature dependence for $y$\,$>$\,0.1 
or for the behavior seen in at $y$\,=\,0.27, 
for which an overall decrease of $\chi$ is predicted. 
%

\subsection{Evolution of ferromagnetic fluctuations and their relation 
to the non-Fermi-liquid behavior}

We now focus on the non-Fermi-liquid behavior around $y$\,$\sim$\,0.20, 
where we observe a low-temperature upturn in the specific heat 
and a clear deviation from $T^{2}$ behavior in the resistivity.
We have described in the previous section that 
we expect a strong enhancement of ferromagnetic fluctuations 
when the Fermi level crosses the vHS---evaluated at $y_{\rm c}$\,=\,0.23 
in the rigid-band-shift model 
\cite{Bergemann_Review}---due to the enhanced DOS 
arising from the shift of the two-dimensional nesting-vector 
$\textit{\textbf{Q}}_{\rm ic}^{\gamma}$ toward $q$\,=\,$0$. 
The enhanced DOS is also expected to affect the specific heat coefficient 
$C_{P}/T$. 
The observed non-Fermi-liquid resistivity exponent of $n$\,=\,1.4 
for $y$\,=\,0.20, 
as shown in the inset of Figure~\ref{Fig4}(a), 
is good agreement with the expectation of the self-consistent renormalization theory with $n$\,=\,$4/3$ for two-dimensional ferromagnetic spin fluctuations 
\cite{Moriya}. 
The essential ingredient here is that the two-dimensional nesting-vector
$\textit{\textbf{Q}}_{\rm ic}^{\gamma}$\,=\,$\textbf{0}$. 
The situation is quite different in the Ti-substituted system 
Sr$_{2}$Ru$_{1-x}$Ti$_{x}$O$_{4}$, 
where at $x_{\rm c}$\,=\,0.025 
the non-Fermi-liquid behavior is observed as well, 
but in that case with a \textit{linear} ($n$\,=\,1) resistivity power-law \cite{Kiku_PRL}.
The origin of this behavior lies in the diverging 
\textit{antiferromagnetic} fluctuations originating from 
the finite (incommensurate) nesting vector 
$\textit{\textbf{Q}}_{\rm ic}^{{\alpha}{\beta}}$ 
between the $\alpha$ and $\beta$ sheets \cite{NFL_Comment}. 
Finally, 
let us discuss the absence of magnetic ordering in 
Sr$_{2-y}$La$_{y}$RuO$_{4}$ around $y$\,=\,$y_{\rm c}$ 
in our current study. 
Deviations from Fermi-liquid behavior are
often observed around a quantum critical point 
in the vicinity of magnetic order \cite{Stewart_Review}.
Indeed, 
the expected divergence of the Lindhard susceptibility 
would strongly promote (ferro)magnetic ordering. 
In Sr$_{2-y}$La$_{y}$RuO$_{4}$ up to $y$\,=\,0.27, 
however, 
we have seen no clear evidence for emergence of magnetic ordering,
although the non-Fermi-liquid behavior is clearly observed around 
$y$\,$\sim$\,0.20, 
indicating the immediate vicinity of the vHS. 
One possibility for the absence of the magnetic ordering seems to be that 
the magnetic correlation length is short as seen 
in the inelastic neutron scattering in pure Sr$_{2}$RuO$_{4}$ 
\cite{Braden_Sr2RuO4};  
the magnetic excitations originating from the $\gamma$ band are widely 
spread around $q$\,$\sim$\,0. 
The short range correlation reminds us of the case of the 
Sr$_{2}$Ru$_{1-x}$Ti$_{x}$O$_{4}$ with $x$\,=\,0.09. 
Elastic neutron scattering measurements 
on this system revealed a clear SDW ordering below 25\,K 
with the nesting vector $\textit{\textbf{Q}}_{\rm ic}^{{\alpha}{\beta}}$. 
However, 
the correlation length at the ordered state is as short as $\sim$\,5\,nm 
\cite{Braden_Ti} and 
no anomaly corresponding to the transition 
is observed in the specific heat \cite{Kiku_PRL}. 
The correlation length in Sr$_{2-y}$La$_{y}$RuO$_{4}$ has not been 
measured yet, 
although 
it will be important to so, 
in order to detect the evolution of the spin fluctuations at $\textit{\textbf{q}}$\,$\sim$\,$\textbf{0}$ by La doping.

%
\section{Summary}

By minimizing the disorder level by substituting La$^{3+}$ 
for Sr$^{2+}$, 
we reported the effect of additional electron-doping in Sr$_{2}$RuO$_{4}$. 
The enhancement of the DOS by additional electrons is well explained 
by a rigid-band model based on the realistic tight-binding calculation. 
The result can be mainly ascribed to the approach of the Fermi energy 
towards a vHS of $\gamma$ band, 
which is positioned at a doping level $y_{\rm c}$\,$\sim$\,0.20, 
49\.meV above the Fermi level in pristine Sr$_2$RuO$_4$. 
The enhancement of the magnetic susceptibility by electron doping 
toward the vHS can be viewed in terms of the shrinking 
of the nesting vector of the $\gamma$ band from 
$\textit{\textbf{Q}}_{\rm ic}^{\gamma}$\,$\sim$\,(0.2$\pi$,\,0.2$\pi$,\,0) at $y$\,=\,0 
to $\textit{\textbf{Q}}_{\rm ic}^{\gamma}$\,$\sim$\,\textbf{0} 
at $y$\,$\sim$\,$y_{\rm c}$ on the basis of a calculation of 
the Lindhard susceptibility.
The non-Fermi-liquid behavior around $y$\,=\,0.20 is explained 
in terms of the two-dimensional ferromagnetic spin fluctuations 
with short range correlation. 
On the other hand, 
Ti substitution induces magnetic ordering associated with 
the nesting between other $\alpha$ and $\beta$ Fermi-surface sheets, 
so appropriate dopants can selectively enhance the spin fluctuations of
different bands in Sr$_{2}$RuO$_{4}$. 
Thus, 
we stress that 
the layered ruthenate Sr$_{2}$RuO$_{4}$ is a prototype 
multi-band metal in which we can understand the physical properties
with a surprising level of precision, 
given the correlated nature of the electronic state in this material. 
%

%
\section*{Acknowledgments} 

The authors thank K.M. Shen, M. Braden, S.R. Julian, 
T. Nomura, Kosaku Yamada, and R. Werner  for useful discussions. 
They also thank H. Fukazawa and H. Yaguchi  
for technical supports and discussions, 
M. Yoshioka for technical supports, 
Y. Shibata, J. Hori, Takashi Suzuki, and Toshizo Fujita 
for EPMA measurements at Hiroshima University. 
This work was supported by the Grant-in-Aid for Scientific Research (S) 
from the Japan Society for Promotion of Science (JSPS), 
by the Grant-in-Aid for Scientific Research on Priority Area 
'Novel Quantum Phenomena in Transition Metal Oxides' 
from the Ministry of Education, Culture, Sports, Science and Technology. 
One of the authors (N.K.) is supported by JSPS Postdoctoral
Fellowships for Research Abroad.
%

%



\begin{thebibliography}{99}

\bibitem{Maeno_Nature}
Y. Maeno, H. Hashimoto, K. Yoshida, S. Nishizaki, T. Fujita, J.G. Bednorz, and F. Lichtenberg, 
Nature \textbf{372}, 532 (1994). 

\bibitem{Andy_Review}
A.P. Mackenzie and Y. Maeno, Rev. Mod. Phys. \textbf{75}, 657 (2003). 

\bibitem{Andy_dHvA}
A.P. Mackenzie, S.R. Julian, A.J. Diver, G.J. McMullan, M.P. Ray, G.G. Lonzarich, 
Y. Maeno, S. Nishizaki, and T. Fujita, Phys. Rev. Lett. \textbf{76}, 3786 (1996). 

\bibitem{Bergemann_PRL}
C. Bergemann, S.R. Julian, A.P. Mackenzie, S. NishiZaki, and Y. Maeno, 
Phys. Rev. Lett. \textbf{84}, 2662 (2000). 

\bibitem{Bergemann_Review}
C. Bergemann, A.P. Mackenzie, S.R. Julian, D. Forsythe, and E. Ohmichi, 
Adv. Phys. \textbf{52}, 639 (2003). 

\bibitem{Damascelli_ARPES}
A. Damascelli, D.H. Lu, K.M. Shen, N.P. Armitage, F. Ronning, D.L. Feng, C. Kim, Z.-X. Shen, 
T. Kimura, Y. Tokura, Z.Q. Mao, and Y. Maeno, Phys. Rev. Lett. \textbf{85}, 5194 (2000). 

\bibitem{Oguchi}
T. Oguchi, Phys. Rev. B. \textbf{51}, 1385(R) (1995). 

\bibitem{Singh_Band}
D.J. Singh, Phys. Rev. B. \textbf{52}, 1358 (1995). 

\bibitem{Maeno_FL}
Y. Maeno, K. Yoshida, H. Hashimoto, S. Nishizaki, S. Ikeda, M. Nohara, T. Fujita, 
A.P. Mackenzie, N.E. Hussey, J.G. Bednorz, and F. Lichtenberg, 
J. Phys. Soc. Jpn. \textbf{66}, 1405 (1997). 

\bibitem{Mathur}
N.D. Mathur, F.M. Grosche, S.R. Julian, I.R. Walker, D.M. Freye, 
R.K.W. Haselwimmer, and G.G. Lonzarich, Nature \textbf{394}, 39 (1998). 

\bibitem{Saxena}
S.S. Saxena, P. Agarwal, K. Ahilan, F.M. Groche, R.K.W. Haselwimmer, 
M.J. Steiner, E. Pugh, I.R. Walker, S.R. Julian, P. Monthoux, G.G. Lonzarich, 
A. Huxley, I. Sheikin, D. Braithwaite, and J. Flouquet, 
Nature \textbf{406}, 587 (2000). 

\bibitem{Aeppli}
G. Aeppli, T.E. Mason, S.M. Hayden, H.A. Mook, and J. Kulda, 
Science \textbf{278}, 1432 (1997). 

\bibitem{Stewart_Review}
G.R. Stewart, Rev. Mod. Phys. \textbf{73}, 797 (2001)

\bibitem{Monthoux}
P. Monthoux and G.G. Lonzarich, Phys. Rev. B. \textbf{59}, 14598 (1999). 

\bibitem{Braden_Sr2RuO4}
M. Braden, Y. Sidis, P. Bourges, P.Pfeuty, J. Kulda, Z. Mao, and Y. Maeno, 
Phys. Rev. B. \textbf{66}, 064522 (2002). 

\bibitem{Sidis}
Y. Sidis, M. Braden, P. Bourges, B. Hennion, S. NishiZaki, Y. Maeno, and Y. Mori, 
Phys. Rev. Lett. \textbf{83}, 3320 (1999). 

\bibitem{Mazin_PRL99}
I.I. Mazin and D.J. Singh, Phys. Rev. Lett. \textbf{82}, 4324 (1999). 

\bibitem{Bergemann_HighField}
C. Bergemann, S.R. Julian, A.P. Mackenzie, A.W. Tyler, D.E. Farrell, 
Y. Maeno, and S. NishiZaki, 
Phisica C \textbf{317-318}, 444 (1999). 

\bibitem{Forsythe}
D. Forsythe, S.R. Julian, C. Bergemann, E. Pugh, M.J. Steiner, 
P.L. Alireza, G.J. McMullan, F. Nakamura, R.K.W. Haselwimmer, I.R. Walker, 
S.S. Saxena, G.G. Lonzarich, A.P. Mackenzie, Z.Q. Mao, and Y. Maeno, 
Phys. Rev. Lett. \textbf{89}, 166402 (2002). 

\bibitem{Braden_Ti}
M. Braden, O. Friedt, Y. Sidis, P. Bourges, M, Minakata, and Y. Maeno, 
Phys. Rev. Lett. \textbf{88}, 197002 (2002).

\bibitem{Kiku_PRL}
N. Kikugawa and Y. Maeno, Phys. Rev. Lett. \textbf{89}, 117001 (2002). 

\bibitem{Ishida_Ti}
K. Ishida, Y. Minami, Y. Kitaoka, S. Nakatsuji, N. Kikugawa, and Y. Maeno, 
Phys. Rev. B \textbf{67}, 214412 (2003). 

\bibitem{Minakata}
M. Minakata and Y. Maeno, Phys. Rev. B \textbf{63}, 180504(R) (2001).  

\bibitem{Klaus}
K. Pucher, J. Hemberger, F. Mayr, V. Fritsch, A. Loidl, E.-W. Scheidt, S, Klimm, R. Horny, S. Horn, S.G. Ebbinghaus, A. Reller, and R.J. Cava, Phys. Rev. B \textbf{65}, 104523 (2002). 

\bibitem{Sigrist}
M. Sigrist, D. Agterberg, A. Furusaki, C. Honerkamp, K.K. Ng, T.M. Rice, 
and M.E. Zhitomirsky, 
Physica C \textbf{317-318}, 134 (1999). 

\bibitem{Nomura}
T. Nomura and K. Yamada, J. Phys. Soc. Jpn. \textbf{69}, 3678 (2000); 
\textit{ibid}. \textbf{71}, 1993 (2002). 

\bibitem{Electron_Comment}
In the ref. \onlinecite{Singh_Band}, 
the additional electrons that Fermi energy approaches the singularity is evaluated as 0.2 electrons/f.u. 

\bibitem{Hashimoto}
H. Hashimoto, Master thesis, Hiroshima University. 

\bibitem{Order_Comment}
Tiny hysteresis between zero-field-cooled and field-cooled process are 
observed at low temperature. 
We concluded that this is is not due to an intrinsic behavior, 
but  probably due to inclusion of ferromagnet SrRuO$_{3}$ 
(or Sr$_{1-y}$La$_{y}$RuO$_{3}$) 
with estimated amount of less than $\sim$\,100\,ppm, 
as low as we could not detect the signal from the X-ray measurements. 
The ferromagnetism in Sr$_{1-y}$La$_{y}$RuO$_{3}$ is 
gradually suppressed with La concentration \cite{Nakatsugawa}. 

\bibitem{Ikeda}
S. Ikeda, Y. Maeno, S. Nakatsuji, M. Kosaka, and Y. Uwatoko, 
Phys. Rev. B \textbf{62}, 6089(R) (2000); 
S. Ikeda, N. Shirakawa, T. Yanagisawa, Y. Yoshida, S. Koikegami, S. Koike, M. Kosaka, and Y. Uwatoko, to appear in J. Phys. Soc. Jpn. (cond-mat/0403074)

\bibitem{Andy_JPSJ}
A.P. Mackenzie, S. Ikeda, Y. Maeno, T. Fujita, S.R. Julian, and G.G. Lonzarich, 
J. Phys. Soc. Jpn. \textbf{67}, 385 (1998). 

\bibitem{Kiku_Rigid}
N. Kikugawa, A.P. Mackenzie, C. Bergemann, R.A. Borzi, S.A. Grigera, and Y. Maeno, 
unpublished. (cond-mat/0403337) 

\bibitem{Kiku_JPSJ}
N. Kikugawa, A.P. Mackenzie, and Y. Maeno, J. Phys. Soc. Jpn. \textbf{72}, 237 (2003). 

\bibitem{Adachi_Comment} 
As seen in Fig.~\ref{Fig5}(a), 
a deviation from the $T$-squared behavior around $y$\,$\sim$\,0.05 is observed. 
%
This might suggest a connection with fluctuations near the critical disorder 
for disappearance of superconductivity 
($y$\,$\sim$\,0.03 in Sr$_{2-y}$La$_{y}$RuO$_{4} $\cite{Kiku_Rigid}) 
as theoretically proposed \cite{Adachi}, 
although the predicted temperature dependence of the fluctuation conductivity 
is very weak. 

\bibitem{Yoshida}
K. Yoshida, Y. Maeno, S. Nishizaki, S. Ikeda, and T. Fujita, J. Low Temp. Phys. \textbf{105}, 1593 (1996). 

\bibitem{Kyle_Comment}
K.M. Shen, private communication. 

\bibitem{Yoshida_Textbook}
For instance, K. Yoshida, \textit{Theory of Magnetism} (Springer-Verlag, 1996). 

\bibitem{Moriya}
M. Hatatani and T. Moriya, J. Phys. Soc. Jpn. \textbf{64}, 3434 (1995).

\bibitem{NFL_Comment}
We note that 
our analysis of $\rho_{ab}$ may be simplistic, 
since resistive exponents in other ruthenates are sensitive to disorder, 
as reported in the ref.~\onlinecite{Capogna}. 

\bibitem{Nakatsugawa}
H. Nakatsugawa, E. Iguchi, and Y. Oohara, J. Phys. Condens. Matter \textbf{14}, 415 (2002). 

\bibitem{Adachi}
H. Adachi and R. Ikeda, J. Phys. Soc. Jpn. \textbf{70}, 2848 (2001). 

\bibitem{Capogna} 
L. Capogna, A.P. Mackenzie, R.S. Perry, S.A. Grigera, L.M. Galvin, 
P. Raychaudhuri, A.J. Schofield, C.S. Alexander, G. Cao, 
S.R. Julian, and Y. Maeno, Phys. Rev. Lett. \textbf{88}, 076602 (2002), 
and references therein. 

\end{thebibliography}
\end{document}